# Bandwidth and Offset Launch Investigations on a 1.4 m Multimode Polymer Spiral Waveguide

Jian Chen[1], Nikos Bamiedakis[1], Richard V. Penty[1], Ian H. White[1],
Petter Westbergh[2], Anders Larsson[2]

[1] University of Cambridge, 9 JJ Thomson Avenue, Cambridge, United Kingdom,
jc791@cam.ac.uk
[2] Chalmers University of Technology, SE-412 96 Göteborg, Sweden

**Abstract:** Bandwidth measurements are conducted on a 1.4 m long spiral polymer multimode waveguide for a SMF and 50/125 μm MMF launch and for different input offsets. The waveguide exhibits a bandwidth of at least 30 GHz for all input types, yielding a bandwidth-length product of at least 42 GHz×m, while no impact is observed on the waveguide performance due to the different spatial input offsets. The results indicate that data transmission at data rates even higher than 25 Gb/s can be achieved over such structures, thereby demonstrating the potential of multimode polymer waveguide technologies in short-reach board-level datacommunication links.

**Introduction:** High bandwidth short-reach optical interconnects have been of particular interest in recent years in high-performance computing and data centre environments due to the inherent drawbacks of traditional electrical interconnects such as high frequency loss, poor immunity to electromagnetic interference and heat dissipation issues.[1,2] Multimode polymer waveguides are potential candidates for use in board-level optical interconnections owing to the large bandwidth, low crosstalk and high density that they can offer. In recent years polymer materials which possess favourable optical, mechanical and thermal properties enabling cost-effective integration on printed circuits boards (PCBs) have been developed.[3,4] The use of large core size optical waveguides further offers relaxed alignment tolerances enabling therefore cost-effective board assembly.[5,6] However, the demand of higher and higher data rates increases for on-board interconnects, the bandwidth limitation of multimode waveguide due to modal dispersion needs to be examined. This paper presents therefore bandwidth studies and offset launch investigations on a 1.4 m long spiral waveguide, which we believe demonstrates a record performance achieving bandwidth-length product of at least 42 GHz×m for this long integrated waveguide.

**Experimental Results:** The 1.4 m long spiral waveguide is fabricated on a 6-inch glass substrate using siloxane polymer materials (Core: Dow Corning® OE-4140 Cured Optical Elastomers; cladding: OE-4141 Cured Optical Elastomer). The core and cladding materials, with bulk refractive indices of 1.52 and 1.50 respectively, are spun onto the substrate directly; and the waveguide is patterned by standard photolithographic methods.[1] The waveguide core has a cross section of $50 \times 20$ μm$^2$ and is sandwiched between the top and cladding layers. A vector network analyser (Agilent 8722ET) is used to measure the $S_{21}$ parameter of the optical link with and without (back-to-back) the waveguide (Fig. 1a and 1b). The frequency response of the waveguide can be obtained from the difference between the recorded frequency response of the waveguide link and the back-to-back link. The waveguide frequency response is investigated under different launch conditions as different mode power distributions at the waveguide input can often result in different levels of multimode dispersion in the guides. A single-mode fibre (SMF) is used to emulate a restricted launch while a "typical" and a quasi-overfilled 50/125 μm multimode fibre (MMF) launch are used to investigate the waveguide performance with a broader mode power distribution at the waveguide input. The "typical" 50/125 μm MMF launch is obtained by directly butt-coupling a cleaved MMF patchcord with the VCSEL source, while the quasi-overfilled 50/125 μm MMF launch is achieved with the use of a mode mixer (Newport FM-1) before the waveguide.

An 850 nm vertical-cavity surface-emitting laser (VCSEL) (bandwidth of ~ 25 GHz)[7] is used as the transmitter, and a photodiode (VIS D30-850M) with bandwidth of 30 GHz is employed as the receiver. The VCSEL is butt-coupled to the input fibre patchcord (either SMF or 50/125 μm MMF), and a cleaved 50/125 μm MMF is used to couple the light output from the waveguide to the photodiode. Index-matching gel is used at the waveguide input and output facets to reduce the scattering losses due to surface roughness. An Agilent NA7766A variable optical attenuator (VOA) is employed to adjust the optical power received by the photodiode. Fig. 1c, 1d and 1e show the obtained frequency response of the waveguide for the SMF, the "typical" and quasi-overfilled 50/125 μm MMF input respectively with the different horizontal input offsets. The presented results are normalised with respect to that of the back-to-back link, showing that the frequency responses remain approximately constant up to the 30 GHz instrumentation limit for all launch conditions.

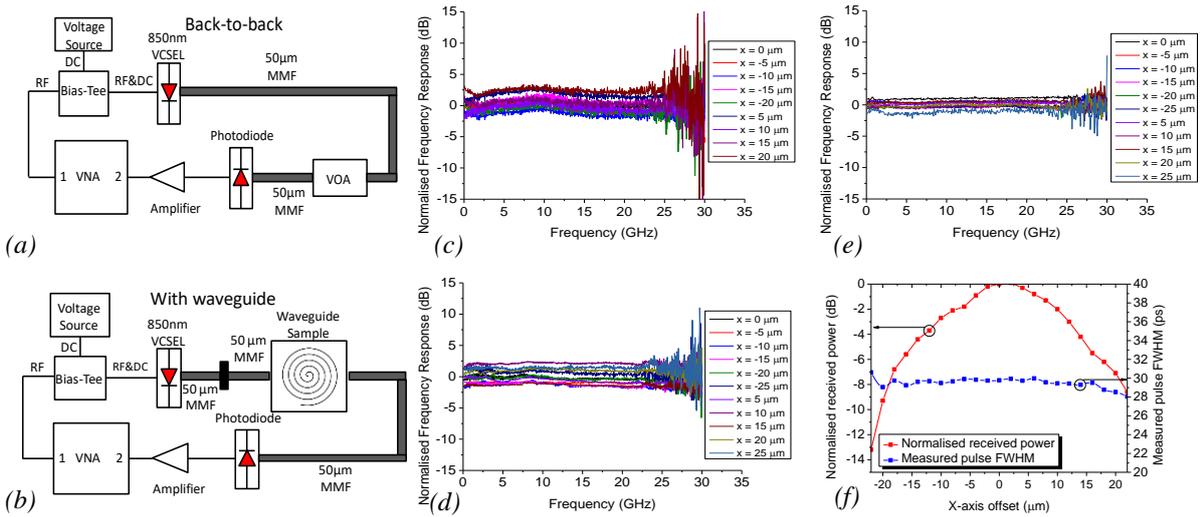

Fig. 1: Experimental set-up of the $S_{21}$ parameter measurement for (a) the back-to-back link and (b) the waveguide link for the 50 μm MMF input. The normalised frequency response of the spiral waveguide for different offsets is shown under (c) a SMF, (d) a typical and (e) a quasi-overfilled 50 μm MMF input; (f) shows the normalised received optical power and FWHM of received signal as a function of the input position.

The dynamic performance of the waveguide is examined for a SMF input to confirm the observations obtained from the $S_{21}$ measurement. A single pulse is generated by modulating the VCSEL with a pulse of duration ~ 23 ps and a repetition rate ~172 MHz. The optical pulse is launched into the spiral waveguide via a SMF and is collected with a 50/125 μm MMF (using a similar set-up to Fig. 1b). The received electrical signal is amplified by a 38 GHz amplifier (SHF 806E), and the FWHM of the received pulse is measured with a digital communication analyser (Agilent 86100A) for the different input positions. The FWHM of the received pulse for the back-to-back link is ~30 ps. As expected from the $S_{21}$ measurement, the FWHM of the received pulse for the waveguide link remains relatively constant for the different input positions, with no significant pulse broadening observed (Fig. 1f). The power received at the waveguide output is also measured for the different input positions (Fig. 1f). 3 dB points are found to be ± 10 μm. Increasing offsets result in larger power coupled into higher order modes which suffer from higher bending losses along the spiral waveguide. The repeatability of the observations is confirmed through multiple measurements. The results indicate that the link's performance does not degrade under different input offsets.

**Conclusions:** Bandwidth measurements have been carried out on a 1.4 m long spiral polymer multimode waveguide. The results indicate that the waveguide exhibits a bandwidth of at least

30 GHz under a large range of launch conditions for all input offsets. This demonstrates the potential of transmitting data rates higher than 25 Gb/s over such structures.

**Acknowledgement:** The authors would like to acknowledge Dow Corning for providing the spiral polymer waveguide samples, EPSRC for supporting the work and IQE Europe for providing the epitaxial material used for the VCSEL fabrication.